\documentclass[twocolumn,epsfig,showpacs,preprintnumbers,superscriptaddress]{revtex4}
\usepackage{graphicx}
\usepackage[floatfix]{epsfig}
\usepackage{dcolumn}
\usepackage{bm}
\def\tc{T$_c$ }

\def\beq{\begin{equation}}
\def\eeq{\end{equation}}
\def\beqn{\begin{eqnarray}}
\def\eeqn{\end{eqnarray}}
\def\eqref#1{Eq.~(\ref{#1}) }

\def\al{\alpha}

\def\eps{\varepsilon}
\def\th{\theta}

\def\ch{\rho}
\def\sp{s}

\def\tp{t_{\perp}}
\newcommand{\hc}{{\rm H.C.}}
\newcommand{\OO}{{\cal O}}
\newcommand{\JJ}{{\cal J}}

\newcommand{\stab}{(See Table~\ref{tab}.) }

\def\domybib{1}

\begin{document}

\title{
The optimal inhomogeneity  for superconductivity
}

\author{E. Arrigoni}
\affiliation{Department of Physics, University of California, Los Angeles,
California 90095, USA}
\affiliation{Institut f\"ur Theoretische Physik und Astrophysik,
Universit\"at W\"urzburg, Am Hubland, D-97074 W\"urzburg, Germany}
\author{S. A. Kivelson}
\affiliation{Department of Physics, University of California, Los Angeles,
California 90095, USA}
\date{\today}

\begin{abstract}

We study the effect of  nonuniform  transverse couplings on a 
 quasi-one dimensional
superconductor.
We show that inhomogeneous  couplings quite generally
increase the superconducting (pairing) gap
relative to the uniform
system, but that beyond an ``optimal'' degree of inhomogeneity, they lead to a
suppression of the tendency to phase coherence. The optimal conditions for
superconductivity
are derived.
We also show that a {\it delocalized}, spin-gapped phase is stable
against weak disorder in a
four-leg-ladder 
with moderate repulsive interactions. 

\end{abstract}


\maketitle

A number of experiments\cite{experiments} including neutron
scattering,
angular resolved
photoemission,
and 
scanning tunnelling microscopy~\cite{stm} 
suggest that many
 high-\tc superconductors show a
highly 
inhomogeneous electronic
structure which often has a
 quasi-one
dimensional nature - ``stripes.''
It is still  unclear
whether inhomogeneities,  and stripes in particular, are 
 an essential feature of high-\tc
superconductivity.

There is 
compelling {\it theoretical} evidence
that quasi-one dimensional systems, such as two-leg Hubbard or $t-J$ ladders
(2LLs), have a strong tendency towards the formation of a spin gap and
substantial superconducting (SC) pair-field correlations~\cite{dagotto}. However, the gap size
and the tendency to superconductivity tends to decrease rapidly with width for ladders with more
legs.~\cite{2ll,sudip}  Due to the phenomenological similarity between 2LLs and two-dimensional
cuprates,  it has been suggested that the pairing mechanism for superconductivity in cuprates may
have a quasi-one-dimensional  origin~\cite{em.ki.za.all}.  However,  strong pairing
is often acompanied by a small superfluid stiffness, especially in
quasi-one-dimensional systems,  so a large spin gap  does not necessarily
imply a high-\tc~\cite{em.ki.95,em.ki.za.all,kive.02}. In other words,
while local pairing    develops easily in a 2LL,
 it is hard for the {\it phase} of the pairs on different weakly
 coupled 2LLs 
to develop coherence.
Another problem with 
quasi-one dimensional superconductors 
is their extreme sensitivity to disorder ~\cite{or.gi.all}. 
Since weak disorder does not destroy superconductivity in two dimensions, it is intuitively clear
 that this sensitivity is mitigated if one increases the number of
 coupled chains.
Taken together, these observations suggest 
there  exists an
intermediate ``optimal'' degree of inhomogeneity~\cite{sudip,er.or.00} which maximizes
 \tc.

In this
Letter, we analyze 
the relation between
inhomogeneity and superconductivity 
in   ``microscopic'' 
inhomogeneous multi-leg ladder Hubbard models
 in which the SC gap arises
solely as a result of {\it repulsive} interactions between electrons.
This allows us to make  {\it quantitative} estimates (at least for weak
coupling) of the optimal degree of inhomogeneity without {\it ad hoc}
assumptions. 
In addition, we show that for a 4-leg ladder  (4LL), there is a broad range of repulsive interactions for
which disorder is irrelevant in the renormalization-group sense.

Specifically,
we have carried out a weak-coupling renormalization-group (RG) analysis 
of a
model (Eq. \ref{h}) of two 2LLs
with a
repulsive on-site interaction $U$ coupled via an interladder hopping
$t$
and with an on-site energy offset $\eps$ between the 2LL.  
We
consider $t$ in the range $1\ge t> 0$; in the units we have
adopted $t=1$ and $\eps=0$ corresponds to a homogeneous 4LL,
while at
$t=0$ there are two (inequivalent) decoupled 2LLs. 
For simplicity, in order to have a unique parameter which determines the degree of inhomogeneity,
$\eps$ is taken to depend on
$t$ as $\epsilon=\epsilon_0(1-t^2)$.  Moreover,
we  have considered both a flat ladder, with open
boundary conditions in the
direction transverse to the chains (OBC), and of a 
cylindrical ladder, with periodic boundary conditions (PBC).

\begin{figure}[htb] 
  \includegraphics[width=7cm]{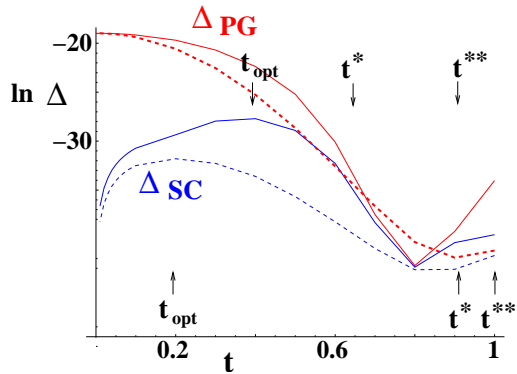}
  \caption{
Logarithm of  $\Delta_{SC}$ and $\Delta_{PG}$
for a 4LL with OBC (solid line) and PBC (dashed)
for $\eps_0=1$, $\nu=1.2$, and $U=1$.
We also indicate the positions of $t_{opt}$, $t^*$, and $t^{**}$ for
OBC (upper row) and PBC (lower).
\vspace*{-.2cm}
}
  \label{gap}
\end{figure}

Our results,
 valid qualitatively in a range electron concentrations per site $\nu$ near but not equal to 1,
are presented in Fig.~\ref{gap}, and can be summarized as follows:      In the parameter range
that we have considered,  there are 4 bands which cross the Fermi energy, so there are
potentially 4 distinct gapless charge ($\ch$) and spin ($\sp$) modes. 
Upon bosonization, these modes are represented by the collective bosonic fields, $\phi_{\ch,a}$ and
$\phi_{\sp,a}$ where $a=1 - 4$ is a band index~\cite{convention}. 
 Phases (that is to say fixed points of the RG flows) are
labelled CnSm according~\cite{2ll} to the number n and m of charge  and spin
modes  that remain gapless in the presence of interactions.  For the entire range of $t$, we
will show that the 4LL is in the maximally gapped C1S0 phase in which all
$4$ spin modes   
and the
$3$ ``relative'' charge modes
are gapped~\cite{kf}. (So long as $\nu$ is irrational, the total charge mode $\phi_{\ch+}\equiv
(1/2)\sum_a \phi_{\ch,a}$, is always gapless\cite{convention}.)  This 
phase
is characterized by  
slowly decaying  
power law SC 
correlations, and can be considered as the ``SC'' phase of
the 4LL.  For a range of $t$, a single, strong-coupling
fixed point governs the physics, and hence all the gaps are ``comparable''~\cite{magn}, in the sense
that they all have the same singular dependence on the interaction strength - this single scale
behavior is remeniscent of an anisotropic BCS superconductor.  In other
ranges, the RG flows pass close to an initial strong coupling fixed
point before finally reaching the C1S0 fixed point. Here, there are two
distinct gap scales, a ``pseudo-gap'' scale, $\Delta_{PG}$ with magnitude
governed by the flow to the initial strong-coupling fixed point, and  a
lower scale, which we call the superconducting gap
$\Delta_{SC}$, governed by the second segment of the flow.  

The crucial result (see Fig.~\ref{gap}) is that 
$\Delta_{SC}$ 
is not a monotonic function
of $t$, and, in particular, it has an absolute maximum at a certain
$t=t_{opt}<1$, i. e. for an {\it inhomogeneous} 4LL.  We find 
three regimes of $t$:  
For 
$ t^{**}\ge t \ge t^* 
$, all the gaps are comparable.
In contrast, two distinct gap scales are found for   $1\ge t > t^{**}$ and $t^* > t > 0$.
The two-gap behavior at small $t$ can be understood as follows:  For $t=0$, one of the 2LLs has an
exponentially larger superconducting gap than the other due to the different Fermi
velocities. 
 For small but non-zero $t$,
$\Delta_{PG}$ is continuously connected to this gap,  i.e. it is the gap associated with the pair of
bands predominantly associated with the more SC 2LL. 
However, for finite $t$, smaller gaps of magnitude $\Delta_{SC}$ are induced in the remaining bands by a
generalized version of the proximity effect~\cite{em.ki.za.all}.

\paragraph{Inhomogeneous 4LL in weak coupling:} We consider 
a Hubbard Hamiltonian for two coupled
2LLs  
\beqn
\label{h}
&&
H=-t_{\parallel}\sum_{y=1}^4\sum_{x,\sigma}\left[c_{x,y,\sigma}^{\dagger}c_{x+1,y,\sigma}+\hc
\right]\\ \nonumber
&& - \sum_{x,\sigma}\tp \bigl[
 c_{x,1,\sigma}^{\dagger}c_{x,2,\sigma}+
c_{x,3,\sigma}^{\dagger}c_{x,4,\sigma}+\hc\bigr] 
\\ \nonumber &&
- \sum_{x,\sigma}\bigl[t
c_{x,2,\sigma}^{\dagger}c_{3,x,\sigma}
+t' c_{x,4,\sigma}^{\dagger}c_{x,1,\sigma}+ \hc
\bigr]\nonumber\\
&& +\sum_{x}\bigl\{ (U/2)\sum_{y=1}^4[\hat n_{x,y}]^2+ \epsilon[\hat
n_{x,1}+\hat n_{x,2}]\bigr\},\nonumber
\eeqn
where, $c_{x,y,\sigma}^{\dagger}$ creates an electron with spin
$\sigma$ on chain $y$ at position $x$ along the chain and  $\hat
n_{x,y}\equiv\sum_{\sigma}c_{x,y,\sigma}^{\dagger}c_{x,y,\sigma}$. 
We choose units in which 
the hopping matrix elements along the ladders,
$t_{\parallel}=2$.   We set 
$\tp=1$ along the rungs of each 2LL, and  a different coupling $t\le \tp$ along
the rungs connecting the two 2LLs where  $t'=0$ corresponds to  OBC and $t'=t$ to  PBC.
In addition, one of the two ladders is shifted in energy by  an
amount
$\epsilon=\epsilon_0(1-t^2)$.
 The electrons
interact via a weak, on-site Hubbard interaction 
$U\ll 2\pi t_{\parallel}$.
We fix 
the number of electrons per site,
 to be close to, but not equal to one.  

\paragraph{ First stage renormalization:}  
It is by now a straightforward procedure~\cite{2ll} to derive the 
 weak-coupling RG equations for the large number  of
distinct running couplings
$g_i$ that define the various low-energy two-particle scattering processes. 
  These equations always  have the form
\beq
d\ g_i /d\
\tau = A^i_{j,l} g_j  g_l +{\cal O}(g^3), 
\label{RG}
\eeq
where 
$\tau$ is the log of the bandwidth (cutoff), and $A^i_{j,l}$
are coefficients which depend on the details of the non-interacting system
under consideration, {\it i.e.} in the present case they depend on $\epsilon_0$,
$t$, and $\nu$.  The other inputs to the calculation are the initial values 
$g_i(\tau=0)= \lambda_j U\ll 1$, where the dimensionless constants $\lambda_j$
also depend on $\epsilon_0$, 
$t$, and $\nu$.  Information about the characteristic emergent energy scales is
obtained by integrating these equations (typically 
numerically) until one or more coupling grows to be of order 1.

To be precise, a characteristic gap scale,
\beq
\label{delta}
\Delta \sim \exp[-\tau^*] = \exp\left[-\frac{\al}{U} +\OO(\log U)\right] \;.
\eeq
is obtained from the  value of
$\tau=\tau^*=\al/U$ at which some set of couplings diverge.  (The $\OO(\log U)$ correction represents the
effect of the neglected cubic terms in powers of $g$.)  For $\tau$ near $\tau^*$, there are generally a
set of strongly divergent couplings, $g_i\sim G_i(\tau^*-\tau)^{-1}$, which all grow to be of order 1 when
$(\tau^*-\tau)\sim 1$, and so define the new ``strong-coupling fixed point.''  
There is also a set of
weakly or non-divergent couplings, $g_i\sim G_iU^{\gamma_i}(\tau^*-\tau)^{-(1-\gamma_i)}$ with
$\gamma_i > 0$, which remain small, $g_i\sim G_i U^{\gamma_i}$, when
$(\tau^*-\tau)\sim 1$~\cite{3c.pla,2ll}.  
What modes are
gapped at this strong-coupling fixed point is then determined by bosonizing the model. The order 1
couplings are considered to be large
 and thus pin (gap)  the appropriate
modes. 
As we will discuss below,
  it may still be necessary to do further analysis to address the fate of the
remaining gapless modes at this  strong-coupling fixed point, and to account for
the effects of the weak residual interactions between 
them~\cite{em.ki.za.all}b.
The coefficient $\al$ in \eqref{delta}, which is the
principle result of  this initial 
analysis, clearly depends on the 
the values of $t$
 $\epsilon_0$, and $\nu$. 

Since  this analysis straightforwardly extends the previous treatments of the 2LL and
4LL~\cite{2ll},
 we will not, here, present the details of the calculation. Some representative
results, for parameters listed in the captions,  are shown in
 Fig.~\ref{gap}.  
For homogeneous $2N$LLs, 
$\al$ increases with the number of
chains $2N$, as pointed out in Ref.~\cite{2ll}, so that 
the gaps
decrease exponentially with $N$. 
In the present case, $\alpha(t)$ must interpolate between the relatively large value for a homogeneous 4LL
when
$t=1$ and the smaller 2LL value when $t=0$ \stab.  By integrating the
one-loop RG equations, we found, to our surprise, that even the dependence of $\alpha$ on $t$ is
not monotonic;  
$\al(t)$ 
 first increases with decreasing $t$ until it reaches a maximum value at $t=t_{min}$, and then decreases
(i. e. the gaps increase exponentially) as $t$ decreases further.

For $t^*<t<t^{**}$  \stab
{\it all} the couplings responsible for gapping the
$3$ charge and $4$ spin  modes diverge in proportion to $(\tau^*-\tau)^{-1}$   at
$\tau=\tau^*$, so that all gaps at this C1S0 fixed point are
``comparable''~\cite{magn}.
On the other hand,
for $t^* \ge t >0$ or $1\ge t>t^{**}$, 
only the couplings 
responsible for 
the  gaps in the spin and relative
charge modes
within two bands
diverge like
$(\tau^*-\tau)^{-1}$.
More specifically, at the resulting (C3S2) fixed point, $\
\Delta_{PG}=\Delta$ characterizes the pinning of
 the spin fields~\cite{convention} $\phi_{\sp,1}, \phi_{\sp,2}$ associated with two of the bands
(labelled, for simplicity, 1 and 2), and of the relative ``superconducting phase,''
 $\th_{\ch,(1-2)}\equiv (\th_{\ch,1}-\th_{\ch,2})/\sqrt{2}$, where $\th_{\ch,a}$ is the field
dual to $\phi_{\ch,a}$.
Behavior of this sort has sometimes been interpreted\cite{2ll,3c.pla} as indicating
a {\it transition}  to a phase with  additional gapless modes, {\it
i. e.}, in this case, 
from a C1S0 phase for $t^{**}>t > t^*$ to a C3S2 phase for $t < t^*$
and $t>t^{**}$. 
  However, we shall 
see\cite{em.ki.za.all}b that
 there is  a crossover, but no phase transition, at $t=t^*$.

\paragraph{ Second-stage renormalization:}
For $t < t^*$, the C3S2 strong-coupling fixed
point to which the weak-coupling flows have carried us 
 can be thought
of as describing a 1D superconductor in bands $b_A=1,2$ 
and two ungapped Luttinger liquids (LL) corresponding to bands $b_B=3,4$.
The fixed-point Hamiltonian thus consists of 
$3$
 gapless charge and $2$ gapless spin modes with a renormalized
ultra-violet cutoff 
$\Delta_{PG}$.
However, 
several residual interactions are left at the end
of the first stage of 
renormalization, and 
it is necessary to carry out a perturbative stability
analysis
of the C3S2 fixed point with respect to these interactions~\cite{em.ki.za.all}.
In the weak-coupling limit, it turns out that
the four singlet Josephson
couplings ($\JJ_{b_A,b_B}$) 
between the bands $b_A$ and $b_B$, {\it i.e.} the amplitudes
to scatter a zero-momentum pair from 
$b_A$ to $b_B$, 
 are the most relevant perturbations and make  the 
C3S2 fixed point unstable. 
Other interactions, such as the interband or intraband scattering processes in
the bands $B$, 
are either irrelevant or marginal in the $U\to 0$ limit.

This can be seen by evaluating
the scaling dimension of the $\JJ_{b_A,b_B}$ ($d_J$, approximately
equal for both $b_B$),
which  can be estimated again using bosonization. 
Specifically, one replaces 
the gapped fields of the two bands $b_A$ with their expectation values, 
 while all  other fields
can be taken as
nearly free, i. e. their (spin and charge) 
LL exponents $K$ can be approximately set to $1$,
 since the remaining couplings are small. 
This procedure (which is standard~\cite{em.ki.za.all}) leads to the estimate
\beq
\label{dj}
d_J \approx 1 + 1/(4 \ K_{\ch A})
\eeq 
where $K_{\ch A}$ is the charge Luttinger exponent of the total charge-mode of bands $b_A$.
Since~\cite{krho}
$d_J\approx 5/4 <2$,  $\JJ_{b_A,b_B}$ are relevant perturbations, and diverge at lower
energy scales within a strong-coupling RG expansion about the C3S2
fixed point, as anticipated.
Again, one can identify the energy scale at which one or more 
$\JJ_{b_A,b_B}$ become of order
$1$ 
with the freezing of the associated modes  and 
with the opening of  corresponding gaps.
Specifically, 
$\JJ_{b_A,b_B}$ 
 open a gap in
the  spin modes ($\phi_{\sp,b_B}$ get locked)
as well as in the relative charge modes
(locking the relative SC phases $(\th_{\ch,1}+\th_{\ch,2}-2\th_{\ch,b_B})/\sqrt{6}$)
resulting in a C1S0 phase.
Since 
all $\JJ_{b_A,b_B}$ are comparable~\cite{magn}
and have the same scaling dimension [\eqref{dj}],
all these gaps 
are comparable as well.
These smaller gaps are thus identified as $\Delta_{SC}$ of the 4LL.
In this 
C1S0 phase, only
the global charge mode $\phi_{\ch+}$
remains ungapped.
The 
 criterion discussed above yields 
\beq
\label{del4}
\Delta_{SC} \approx \Delta_{PG} \ \JJ_{AB}^{*\ 1/(2-d_J)}  \;.
\eeq
Here,   $\JJ_{AB}^*$ is the value  of the  
 residual Josephson couplings $\JJ_{b_A,b_B}$ left at the
end of the first stage of renormalization, i. e. when
$(\tau^*-\tau)\sim 1$.
The numerical asymptotic analysis around the singular point $\tau^*$
yields a weak divergence of the $\JJ_{b_A,b_B}$
 with a 
unique exponent 
$\gamma_{AB}\approx 0.06$. Thus, the
 $\JJ_{b_A,b_B}$ attain a value
\beq
\JJ_{b_A,b_B}^*\approx \JJ_{A,B}^*\sim U^{\gamma_{AB}}G_{AB} \;,
\eeq
which is small but finite for small $U$.
(It is interesting that $\gamma_{AB}\approx 0.06$ is independent of $t$ \stab,
although clearly $ G_{AB}$ vanishes
for
$t\to 0$.)

\begin{table}[htb]
$
\begin{array}{l||l|l|l||l|l|l||l||l}
        & t^{*} & t^{**} & t_{min} & \al_0 & \al_1 & \al(t=1) & \gamma_{AB} 
                                     &    G_{AB}^{(2)} \\
\hline
{\rm OBC}  &  0.65   &  0.91    &  0.81    & 20.1  & 16.8  & 35.1 & 0.06    
                                     & 0.05                   \\
{\rm PBC}  &  0.92   &  1.0     &  0.91    & 20.1  & 38.9  & 41.8 & 0.06
                                     & 0.01 \\   
\end{array}
$
\caption{\label{tab}
Values of the parameters 
discussed in the text for a 4LL with OBC and PBC, and for $\nu=1.2$,
and $\eps_0=1$.
\vspace*{-.4cm}
}
\end{table}

The optimal $t$ is not given by $t=t^*$.  
Although the ratio
$\Delta_{SC}
/\Delta_{PG}$ is a rapidly decreasing function of decreasing $t$ for $t<t^*$,
$\Delta_{SC}$ itself  still, in general, initially increases 
due to the exponential increase of $\Delta_{PG}$.
Anticipating the fact that the optimum value for $t^2$ is 
of the order $U$, we can determine it by 
expanding
$\al$ for small $t$ as $\al(t) = \al_0+\al_1 t^2 +
\OO(t^4)$, with $\al_1$ given in Table~\ref{tab}.
Similarly, 
$ G_{AB} =  G_{AB}^{(2)}
\ t^2 + \OO(t^4)$.
Then from \eqref{del4} and 
from the expression for $\JJ_{A,B}^*$
\beq
\label{delsc}
{\Delta_{SC}(t)} \approx {\Delta_{PG}(t=0)} \ \left[ U^{\gamma_{AB}} G_{AB}^{(2)} t^2 \right]^{x_J} \ e^{
-\al_1
  t^2/U}  \;,
\eeq
where $x_J\equiv1/(2-d_J)$.
It is now
straightforward to determine the value of  $t$ which maximizes
$\Delta_{SC}$:
\beq
\label{t1opt}
t_{opt}^2 = \frac{U }{(2-d_J)\al_1} + \OO(U^2) 
\eeq

\paragraph{Quantitative considerations:}  
The various quantities describing the $t$ dependence of
the gaps
discussed here are reported in
Table~\ref{tab} for $\nu=1.2$ and $\epsilon_0=1$.
The curves in Fig.~\ref{gap} have been 
computed as follows.  We have taken initial values of the coupling constants
corresponding to $U=1$ and have  integrated \eqref{RG} numerically 
for different values of $t$
until the largest
of the couplings (in modulus) has reached the
value $1$.
 This determines $\tau^*(t)$, and where needed,
$\JJ_{AB}^*$.
The gaps 
are then evaluated by using
\eqref{delta}, and \eqref{del4} with $d_J=5/4$, respectively.

A different situation occurs when $\epsilon_0=0$.
In this case, the wave functions 
of the
$U=t=0$ part of the Hamiltonian 
are equally distributed on the two 2LL for any 
nonvanishing $t$.
As a consequence, the dependence of the gaps on $t$ obtained by the
method
described above is very weak (for PBC there is no dependence at
all).
 In this case, one should use a different method in which 
couplings with crystal momentum  not exactly conserved at the Fermi
points have to be included in the first stage of the
renormalization.
Qualitatively, the results are similar to the $\epsilon_0\not=0$
case, i. e., the SC gap increases
with increasing
inhomogeneity, there is an optimal value  $t=t_{opt}$, and a
``pseudogap'' behavior. However, in this case $t_{opt}$ is much
smaller
and of the order of the 2LL gap $\Delta_{PG}(t=0)$~\cite{elsewhere}.

\paragraph{The effects of disorder.}
 Orignac and Giamarchi~\cite{or.gi.all} have shown that 
disorder is a relevant perturbation (i.e. leads to localization)
in the C1S0 phase of the  2LL 
unless there are strong enough {\it attractive} interactions such that
the Luttinger exponent, $K_{\ch+}$, 
associated with the gapless total charge mode is greater than $3/2$~\cite{krho}. 
We 
now consider the effect of a single-particle disorder potential 
in the C1S0 phase of the  4LL.

Neglecting 
forward scattering terms, which turn out not too be crucial here,
 one can write this potential as
\beq
\label{hpot}
H_{dis} = \int \ d \ x 
\sum_{b, b'=1}^4 \hat v_{b,b'}(x)e^{i[k_{F,b}+k_{F,b'}]x} + H.C. 
\eeq
Here, the operator $\hat v_{b,b'}(x)$ 
scatters an electron  from band $b$ to band $b'$ and from left to right.
$\hat v_{b,b'}$ is straightforwardly expressed in terms of the bosonic
fields.  
Since in
the C1S0 phase, all spin-fluctuations are gapped, to study the low energy consequences
of weak disorder we can replace all factors of the form $e^{i\alpha
  \phi_{\sp,b}}$  by its
constant expectation value.   
The remaining dependence on the charge and dual spin fields is
\beq
\label{hint}
\hat v_{b,b'}(x) 
\sim  \frac{\xi_{b,b'}(x)}{2 \pi a}
e^{ i  \frac{\phi_{\ch,b}+ \phi_{\ch,b'}}{\sqrt{2}}} \cos\left( 
                   \frac{\th_{\sp,b}- \th_{\sp,b'}}{\sqrt{2}} + \eta \right)  \;
\eeq
where the $x$ dependence of the bosonic fields is left implicit, and
$\eta$ is a $x$-independent phase.
 The $\xi_{b,b'}(x)$ are the random potentials 
which, for the sake of
 definiteness, can be taken to obey a Gaussian distribution
$\overline{\xi_{b_1,b_2}(x) \xi_{b_3,b_4}(y)^*} = W_{b_1,b_2|b_3,b_4} D
\delta(x-y)$, where $D$ is the disorder strength (proportional to the
inverse of the mean-free path), and
 $W$ are numbers of order unity that depend upon 
details of the band structure.
The precise form of the
 $W$
as well as the value of $\eta$
 are
 not important for the following analysis.

Now we get to the nub of the problem.  The only low energy
fluctuations in the C1S0 phase involve the total 
charge mode,
$\phi_{\ch+}$.  Since the spin-fields are pinned, the dual spin fields
are wildly fluctuating, so that any 
operator that depends on $e^{i\alpha\th_{\sp,b}}$ has exponentially
falling correlations, and is hence 
irrelevant in the RG sense.  Similarly, since the relative superconducting phases,
$\th_{\ch,b}-\th_{\ch,b'}$, of the various bands are pinned, any
operator with a factor 
$e^{i\alpha[\phi_{\ch,b}-\phi_{\ch,b'}]}$ is similarly irrelevant.
Therefore, both terms in \eqref{hint} are 
irrelevant at low
temperatures.

This does not mean disorder is irrelevant.
As discussed in Refs.~\onlinecite{or.gi.96,ar.br.96},
additional terms get generated in early stages of the RG flow as the
gapped fields get integrated out. 
It is straightforward to show that the 
 most relevant term is obtained only at fourth order in $\hat v$,
and has the form
\beq
\label{heff}
H_{eff} = \int \ d \ x  \ \frac{\xi_{eff}(x)}{2 \pi a} 
            e^{i\ \sqrt{8} \ \phi_{\ch+}} + \hc \;,
\eeq
where $\xi_{eff}$ is an effective disorder potential,
which is
 again Gaussian distributed:
$\overline{\xi_{eff}(x) \xi_{eff}(y)^*} =  D_{eff} \delta(x-y)$, where
$D_{eff} \sim D (D/v_F^2)^3$.

For weak enough bare disorder~\cite{weaker}, 
and for energies below the SC gap $\Delta_{SC}$,
the RG
 flow for the renormalized disorder strength 
$D_{eff}$ associated with \eqref{heff} 
can be easily derived in the usual way~\cite{or.gi.all}. The RG
equation reads
\beq
\frac{d\ D_{eff}}{d\ \tau} = D_{eff} (3-4 K_{\ch+}) \;,
\eeq
 i. e.
disorder is irrelevant for $K_{\ch+}>3/4$.
This results is very important, as  it signals the presence of  
a localized-delocalized
transition in a quasi-one-dimensional system {\it for purely repulsive
  interactions}~\cite{krho}.
Moreover, the fact that delocalization is present in weak coupling,
i. e. within the range of validity of the present RG procedure,
makes the 4LL one of few low-dimensional models in which one can show the occurrence
of delocalized states in a controlled manner.
This result is valid provided the disorder does not destroy the gaps,
i. e. $D \ll v_F \Delta_{SC}$~\cite{weaker}.
\paragraph{Comments and Speculations.}  The present results were derived for small $U$, where all
the emergent energy scales are exponentially small.  In many cases, the
same physics, and in particular the existence of an optimal degree of inhomogeneity, can be
shown\cite{em.ki.za.all,elsewhere} to apply in the strong-coupling limit, as well, where the gap
scales are a significant fraction of the exchange interaction, $J$.  However, quantitative
results in this regime generally require some input from numerical experiments.

We note that
the present results for the 4LL 
are qualitatively similar to what happens
in  high-\tc superconductors such as
La$_{2-x}$Sr$_x$CuO$_4$~\cite{neutron,ya.le.98}.
In the overdoped region the system is relatively homogeneous,  
both the pairing and SC scales are relatively small and roughly
equal. With decreasing doped hole 
concentratin, $x$, stripe correlations become more prominent and the
interstripe distance increases, 
i. e. the system becomes more effectively inhomogeneous. 
At the same time, the pseudo-gap scale increases monotonically
with decreasing $x$, while the SC $T_c$ has a maximum at a
``optimum'' $x$ and then decreases.
Besides High-\tc superconductors,
 the effects studied here
 could be relevant for  quasi-one-dimensional systems,
such as
 quantum wires or carbon nanotubes, in
which more than one channel cross the Fermi surface.
Finally, this idea could,
 possibly, give direction to the search for better
superconductors~\cite{kive.02}.

We would like to acknowledge useful discussions with T. Giamarchi,
D. J. Scalapino, and O. Zachar.
EA was supported, in part, by the DFG 
via a Heisenberg fellowship (AR 324/3-1) and by DOE grant
n. DE-FG03-00ER45798.  
SAK was suppported, in part, by NSF grant \#DMR 01-10329 at
UCLA.

\ifnum\domybib>0
\bibliographystyle{prsty} 
\bibliography{references_database,footnotes}
\else
\bibliographystyle{prsty} 
\bibliography{references}

\begin{thebibliography}{10}
\bibitem{convention}
We adopt a normalization convention such that $\rho_b(x)=- \de_x
  \phi_{\ch,b}/\pi$ is the long wave-length charge density associated with band
  $b$.
\bibitem{kf}
We consider only parameters for which the four Fermi momenta $k_{F,a}$ are
  mutually incommensurate.
\bibitem{magn}
We define two quantities to be ``comparable'' if their ratio remains finite in
  the $U\to 0$ limit.
\bibitem{krho}
In \eqref{dj}, $K_{\ch(1+2)}$ is the LL exponent associated with
  $\th_{\ch(1+2)}$. One expects that for weak repulsive interactions and not
  too different Fermi velocities both $K_{\ch (1+2)}$, and $K_{\ch+} \approx
  1$, similarly to the two-chain case~\cite{2ll,ar.br.96}.
\bibitem{weaker}
Actually, one can show~\cite{elsewhere} that in the ``pseudogap'' regime a
  weaker condition is sufficient, namely $[D/(v_F \Delta_{PG})]^2
  (\Delta_{SC}/\Delta_{PG}) \ll 1$.
\end{thebibliography}
\fi

\end{document}